\documentclass{article}
\usepackage{amssymb}
\usepackage{amsmath,amsfonts,amsthm}
\usepackage[affil-it]{authblk}
\usepackage{amstext}
\usepackage{amsgen}
\usepackage{amsbsy}
\usepackage{amsopn}
\usepackage{graphicx}
\usepackage{epsfig}
\usepackage[ampersand]{easylist}

\newtheorem{theorema}{Theorem}
\newtheorem{algorithm}[theorema]{Algorithm}
\newtheorem{prop}{Theorem}
\newtheorem{proposition}[prop]{Proposition}
\newcommand{\Prob}{$\mathcal{P}$}

\newcommand{\SSAT}{{SSAT}$(n,m)$}
\def\IMAGESPATH{.}

\input{tcilatex}
\begin{document}

\title{A novel algorithm for solving the Decision Boolean Satisfiability
Problem without algebra}
\author{Carlos Barr\'{o}n-Romero \thanks{%
Universidad Aut\'{o}noma Metropolitana, Unidad Azcapotzalco, Av. San Pablo
No. 180, Col. Reynosa Tamaulipas, C.P. 02200. MEXICO. }}
\date{April 27, 2016}
\maketitle

\begin{abstract}
This paper depicts an algorithm for solving the Decision Boolean
Satisfiability Problem using the binary numerical properties of a
Special Decision Satisfiability Problem, parallel execution,
object oriented, and short termination. The two operations:
expansion and simplification are used to explains why using
algebra grows the resolution steps. It is proved that its
complexity has an upper bound of $2^{n-1}$ where $n$ is the number
of logical variables of the given problem.
\end{abstract}

\hyphenation{corres-pond}

% \begin{keywords}
% \textbf{Key words.}:
Algorithms, Complexity, SAT, NP, CNF. % \bigskip
% \end{keywords}

% \begin{AMS}
% \textbf{AMS subject classifications.}
68Q10, 68Q12,68Q19,68Q25. % \end{AMS}

\pagestyle{myheadings} \thispagestyle{plain}
\markboth{Carlos
 Barr\'{o}n-Romero}{Lower bound for SAT}

%%%%%%%%%%%%%%%%%%%%%%%%%%%%%%%%%%%%%%%%%%%%%%%%%%%%%%%%%%%%%%%%%%%%%%%%%%%%
%%%%%%%%%%%%%%%%%%%%%%%%%%%%%%%%%%%%%%%%%%%%%%%%%%%%%%%%%%%%%%%%%%%%%%%%%%%%

\section{Introduction}

%%%%%%%%%%%%%%%%%%%%%%%%%%%%%%%%%%%%%%%%%%%%%%%%%%%%%%%%%%%%%%%%%%%%%%%%%%%%
%%%%%%%%%%%%%%%%%%%%%%%%%%%%%%%%%%%%%%%%%%%%%%%%%%%%%%%%%%%%%%%%%%%%%%%%%%%%
%%%%%%%%%%%%%%%%%%%%%%%%%%%%%%%%%%%%%%%%%%%%%%%%%%%%%%%%%%%%%%%%%%%%%%%%%%%%
%%%%%%%%%%%%%%%%%%%%%%%%%%%%%%%%%%%%%%%%%%%%%%%%%%%%%%%%%%%%%%%%%%%%%%%%%%%%

This paper focuses in solving the classical Decision Boolean
Satisfiability Problem (SAT) using the results
in~\cite{arXiv:Barron2016} for an special case of the Decision
Boolean Satisfiability Problem, named Simple SAT (SSAT).
In~\cite{arXiv:Barron2010}, the chapter 6 depicts Reducibility.
This term means the ability to solve a problem by finding and
solving simple subproblems. This idea is used here between SAT and
SSAT.

I focused to the CNF version of SAT, which it is justified for the
logical equivalence and the abroad literature. Talking about SAT is
immediately related to NP Class of Problem and its
algothms~\cite{Pudlak1998,Zhang:2001:ECD:603095.603153,
Zhang2002,TOVEY198485,coe:Woeginger2003,pnp:page,
Fortnow:2009:SPV:1562164.1562186,Cook:2000,
Gutfreund:2007:NLH:1341675.1341676}.

The dominant characteristic of SSAT is that all its formulas have the same
number of logical variables. Briefly, the main results in~\cite%
{arXiv:Barron2016} are the lower upper bound $2^{n-1}$ of the
SSAT's algorithms for an extreme case with one or none solution,
and such algorithms realize one lecture of the SSAT's or one
exploration of the research space $[0,2^{n}-1]$ and they are
numerical without algebra. Therefore the complexity for
SSAT$(n,m)$ (where $n$ is the number of boolean variables and $m$
the number of rows) is that the existence of the solution is
$\mathbf{O}(1)$ when $m<2^{n}$ and it is bounded by
$\mathbf{O}2^{n-1})$
for $m\geq 2^{n}$ or $m\gg 2^{n}.$ On the other hand, in~\cite%
{arXiv:Barron2015b} it is depicted that for SSAT does not exist an efficient
algorithm for solving it. This shortly, because the extreme problem SSAT$%
(n,m)$ with one or none solution and repeated rows, $(m\gg 2^{n})$ requieres
one exploration of the rows. The extreme problem SSAT$(n,m)$ can be builded
selecting randomly one number or none in $[0,2{n}-1]$. Then the only way to
infers the solution is verifying that the SSAT$(n,m)$'s rows corresponds to $%
2^{n}-1$ of $2^{n}$ different formulas. On the other hand, the extreme
problem SSAT$(n,m)$ corresponds to a very easy SAT$(n,m^{\prime })$. This
article analyzes how to solve SAT using properties of SSAT and without using
algebra (\cite{arXiv:Barron2015b}, and ~\cite{arXiv:Barron2016}).

The importance of the SAT's complexity is that SAT is a complete
NP problem, and if it has an efficient algorithm then any NP
problem can be solved efficiently. Here, a
carefully resume of the preeminent properties of the SSAT and SAT
are depicted and used to build an special algorithm for solving
any SAT.

Some propositions of my previous works over the  SAT, and SSAT in~%
\cite{arXiv:Barron2005}, ~\cite{arXiv:Barron2010}, ~\cite{arXiv:Barron2015b}%
, and ~\cite{arXiv:Barron2016} are repeated for making this article
self-content. It is worth to mention that the binary numerical approach
depicted here is for solving any SAT with one lecture of the SAT's rows with
the upper bound $2^{n-1}$ steps.

Section~\ref{sc:nota_conve} contains basic definition, concepts
and conventions used. The section~\ref{sc:PropForSolSAT} depicts
the propositions that they characterize necessary and sufficient
conditions for building an algorithm for SAT. The next
section~\ref{sc:algthmsSAT} depicts the algorithms of the parallel
algorithm for solving SAT. It is proved that lower upper bound of
the iterations for solving SAT is $2^n$, where $n$ is the number
of logical variables. Section~\ref{sc:compleForSSAT} discusses and
justifies why the strategies of using SSAT's properties is less
complex for solving SAT than algebra procedures. Particularly, two
operations, expansion and simplification are depicted to connect
SSAT and SAT. The last section depicts my conclusions and future
work.

%%%%%%%%%%%%%%%%%%%%%%%%%%%%%%%%%%%%%%%%%%%%%%%%%%%%%%%%%%%%%%%%%%%%%%%%%%%%
%%%%%%%%%%%%%%%%%%%%%%%%%%%%%%%%%%%%%%%%%%%%%%%%%%%%%%%%%%%%%%%%%%%%%%%%%%%%

\section{Notation and conventions}

~\label{sc:nota_conve}
%%%%%%%%%%%%%%%%%%%%%%%%%%%%%%%%%%%%%%%%%%%%%%%%%%%%%%%%%%%%%%%%%%%%%%%%%%%%
%%%%%%%%%%%%%%%%%%%%%%%%%%%%%%%%%%%%%%%%%%%%%%%%%%%%%%%%%%%%%%%%%%%%%%%%%%%%
%%%%%%%%%%%%%%%%%%%%%%%%%%%%%%%%%%%%%%%%%%%%%%%%%%%%%%%%%%%%%%%%%%%%%%%%%%%%
%%%%%%%%%%%%%%%%%%%%%%%%%%%%%%%%%%%%%%%%%%%%%%%%%%%%%%%%%%%%%%%%%%%%%%%%%%%%

A boolean variable only takes the values: $0$ (false) or $1$
(true). The logical operators are \textbf{not}: $\overline{x};$
\textbf{and}: $\wedge ,$ and \textbf{or}: $\vee .$

Hereafter, $\Sigma =\left\{0,1\right\}$ is the corresponding binary
alphabet. A binary string , $x$ $\in$ $\Sigma^n$ is mapped to its
corresponding binary number in $[0,2^n-1]$ and reciprocally. Moreover, $%
x_{n-1} \vee \overline{x}_{n-2}\ldots x_1 \vee x_0$ corresponds to the
binary string $b_{n-1}b_{n-2}\ldots b_{1} b_0$ where $b_{i}=\left\{
\begin{array}{cc}
0 & \text{if }\overline{x}_{i}, \\
1 & \text{otherwise.}%
\end{array}
\right.$

Such translation has no cost, if $x$ and $\overline{x}$ are represented in
the ASCII code by a (097) and \^{a} (226).

The table for the translation is $C$
\begin{tabular}{l|l|}
index & $C$ \\
$\vdots $ & $\vdots $ \\ \hline
097 & 1 \\ \hline
$\vdots $ & $\vdots $ \\ \hline
226 & 0 \\ \hline
$\vdots $ & $\vdots $%
\end{tabular}

The clause $(x_{n-1} \vee \overline{x}_{n-2}\ldots x_1 \vee x_0)$
is written as a$_{n-1}$ \^{a}$_{n-2}$ $\ldots$ a$_1$ a$_0$.

The translation is $C[c_{n-1}]C[c_{n-2}]\ldots C[c_{1}]C[c_{0}]$ where $c_i$
is the logical variable a$_i$ or \^{a}$_i$ in the position $i$. Hereafter,
the SAT's formulas are translated to its corresponding binary string.

A SAT$(n,m)$ problem consists to answer if a system of $m$ clauses
or row boolean formulas in conjunctive normal form over $n$
boolean variables has an assignation of logical values such the
system of formulas are true.

The system of formulas is represented as a matrix, where each
clause or row formula corresponds to a disjunctive clause. By
example, let SAT$(4,3)$ be

\begin{equation*}
\begin{array}{ccccc}
\ \  & (x_{3} & \vee \ \overline{x}_{2} &  & \vee \ x_{0}) \\
\wedge &  & ( x_{2} & \vee \ x_{1} & \vee \ x_{0}) \\
\wedge & \  & ( \overline{x}_{2} & \vee \ x_{1} & \vee \ x_{0}) \\
\wedge & (x_3 &  &  & \ \vee \ \overline{x}_{0}).%
\end{array}%
\end{equation*}

This problem is satisfactory. The assignation
$x_{0}=1,x_{1}=0,x_{2}=1,$ and $x_{3}=1$ is a solution, as it is
depicting by substituting the boolean values:
\begin{equation*}
\begin{array}{ccccc}
\ \  & (1 & \vee \ 0 &  & \vee \ 1) \\
\wedge &  & ( 1 & \vee \ 0 & \vee\ 1) \\
\wedge &  & ( 0 & \vee \ 0 & \vee \ 1) \\
\wedge & (1 &  &  & \vee \ 0)%
\end{array}
\equiv 1.
\end{equation*}

A simple of SAT [SSAT$(n,m)$] is a SAT$\left( n,m\right) $ with
the requirement that its rows have the same number of boolean
variables in a given order. The variables' order for SAT or SSAT
do not imply to rewrite the problem, it has a constant cost that
can be assumed at the lecture of each formula.

Hereafter, the boolean variables of SAT are identified by $x$ with
subindexes
from $[0,n-1]$, i.e., $x_{n-1},x_{n-2},\ldots ,x_{1},x_{0}$. For the set $%
X=\{x_{n-1},x_{n-2},\ldots ,x_{1},$ $x_{0}\},$ as in~\cite{arXiv:Barron2010}%
, Prop. 4. an enumeration for address identification of any subset
of logical variables is as follows:
\begin{equation*}
\begin{array}{ccr}
\text{Set of Boolean Variables} &  & \mathbb{N} \\
\{\} & \leftrightarrow & 0 \\
\{x_{0}\} & \leftrightarrow & 1=\binom{n}{0} \\
\{x_{1}\} & \leftrightarrow & 2=\binom{n}{0}+1 \\
\vdots & \vdots & \vdots \\
\{x_{1},x_{0}\} & \leftrightarrow & l_0=\sum_{k=0}^{1}\binom{n}{k} \\
\{x_{2},x_{0}\} & \leftrightarrow & l_1=\sum_{k=0}^{1}\binom{n}{k}+1 \\
\{x_{2},x_{1}\} & \leftrightarrow & l_2=\sum_{k=0}^{1}\binom{n}{k}+2 \\
\vdots & \vdots & \vdots \\
X=\{x_{n-1},x_{n-2},\ldots ,x_{1},x_{0}\} & \leftrightarrow &
2^{n}-1 = \sum_{k=0}^{n-1}\binom{n}{k}%
\end{array}%
\end{equation*}

The function $IV:2^{X}\rightarrow \lbrack 0,2^{n}-1]$ gives an unique
identification as a natural number for any subset of boolean variables
of $X$. It could be possible to define an address polynomial for $IV$,
but for the moment the next algorithm can help for building the previos
correspondence.

\begin{algorithm}
~\label{alg:IdxSet} \textbf{Input:} $x=\{x_k,x_{k-1}, \ldots,x_1,x_0\}$: Set
of logical variables.

\textbf{Output:} $ix$: integer; // the unique index in $[0,2^n-1]$ for the
set $x$.

\textbf{Variables in memory}: $base$: integer; $v,t$: variable set treated
as a number of base $\{n-1,n-2,\ldots,1,0\}$

\vspace{-1mm}\noindent\hrulefill
\begin{easylist}
\renewcommand{\labelitemi}{\ }
 \setlength\itemsep{-0.2mm}

\item $ix = 0;$

\item $k = |x|;$ // where $|\cdot|$ is the cardinality function.

\item \textbf{if} $k$ \textbf{equals} $0$ \textbf{then}

\item \hspace{0.5cm} \textbf{output}: "$ix$.";

\item \hspace{0.5cm} \textbf{stop};

\item \textbf{end if}

\item $base = 0$;

\textbf{for} $j=0,k-1$ \textbf{do}

\item \hspace{0.5cm} $base = base + \binom{n}{j};$ // $\binom{\cdot}{\cdot}$
binomial coefficient

\item \textbf{end do}

\item \textbf{build} $v=v_k,\ldots,v_0$; // the lower set of variables in
order of size $k$

\item \textbf{While} $(x < v)$ \textbf{do}.

\item \hspace{0.5cm} $t=v;$

\item \hspace{0.5cm} \textbf{repeat}

\item \hspace{0.5cm} \hspace{0.5cm} $t=t+1;$

\item \hspace{0.5cm} \hspace{0.5cm} \textbf{if } variables in $t$ different
and in descending order \textbf{then}

\item \hspace{0.5cm}\hspace{0.5cm}\hspace{0.5cm} \textbf{exit}

\item \hspace{0.5cm}\hspace{0.5cm} \textbf{end if}

\item \hspace{0.5cm} \textbf{until} false;

\item \hspace{0.5cm} $v=t$;

\item \hspace{0.5cm} $ix=ix+1$;.

\item \textbf{end while}

\item \textbf{output}: "$base+ix$.";

\item \textbf{stop};
\end{easylist}
\vspace{-2mm}\noindent\hrulefill
\end{algorithm}

Giving any SAT$(n,m)$ and $r$ any clause of it, then $r$ can be
associated to
an unique and appropriate SSAT$(IV(r))$ using the algorithm~\ref%
{alg:IdxSet}, where $r$ represents the set of boolean variables in the
clause $r$ with the convention that they subindex are in descending
order.

The cross-join operator ($\times\theta$) corresponds with two
operations: a cross product and the natural join or ($\theta$)
join. the $theta$ join is like the relational data base natural
join operation. It is used between two set of variables $r$ and
$r^\prime$ and the solutions of SSAT$(\cdot).S$ as follow:

SSAT$(IV(r))$ $\times \theta $ SSAT$(IV(r^{\prime }))=$

\begin{enumerate}
\item \textbf{if} $r \cap r^{\prime } =\emptyset$ \textbf{then}
SSAT$(IV(r)).S\, \times\,$ SSAT$(IV(r^{\prime })).S$.

\item \textbf{if} $r \cap r^{\prime} \ne \emptyset$  and there are
common values between

SSAT$(IV(r)).S$ and SSAT$(IV(r^{\prime })).S$ for the variables in
$r \cap r^{\prime}$ \textbf{then}

$\text{SSAT}(IV(r)).S \,  \theta _{r \cap r^{\prime}} \,
SSAT(IV(r^{\prime })).S$

\item \textbf{if} $r \cap r^{\prime } \ne \emptyset $ and there
are not a common values between

SSAT$(IV(r)).S$  and SSAT$(IV(r^{\prime })).S$ for the variables
in $r \cap r^{\prime}$ \textbf{then} $\emptyset.$
\end{enumerate}

In the case 1 and 2, SSAT$(IV(r)$ and SSAT$(IV(r^{\prime })$ are
compatibles. In the case 3) they are incompatibles, i.e., there is
not a satisfactory assignation for both.

An example of the case 1) is the following $\varphi_1=$SAT$\left(
4,5\right) $
\begin{equation*}
\begin{array}{ccc}
\ \  & (x_{3}\vee \overline{x}_{2}) &  \\
\wedge & (x_{3}\vee x_{2}) &  \\
\wedge & (\overline{x}_{3}\vee x_{2}) &  \\
\wedge & \  & (\overline{x}_{1}\vee \overline{x}_{0}) \\
\wedge & \text{ } & (x_{1}\vee x_{0})%
\end{array}
.
\end{equation*}

A SSAT$\left( 2,3\right) $ is the three first clauses of $\varphi_1$, its solution is $%
\left[
\begin{array}{cc}
x_{3} & x_{2} \\
1 & 1%
\end{array}
\right] .$

A SSAT$\left( 2,2\right) $ is the last two clauses of $\varphi_1$, its solutions are $%
\left[
\begin{array}{cc}
x_{1} & x_{0} \\
0 & 1 \\
1 & 0%
\end{array}
\right] .$

Then the solutions of SAT$\left( 4,5\right) $ are $\left\{ \left(
1,1\right) \right\} \times \left\{ \left( 0,1\right) ,\left(
1,0\right) \right\}$ $=$

 $\left[
\begin{array}{cccc}
x_{3} & x_{2} & x_{1} & x_{0} \\
1 & 1 & 0 & 1 \\
1 & 1 & 1 & 0%
\end{array}
\right] .$

An example of the case 2) is the following $\varphi_2=$SAT$\left(
4,5\right) $

\begin{equation*}
\begin{array}{ccc}
\ \  & (x_{3}\vee \overline{x}_{2}) &  \\
\wedge & (\overline{x}_{3}\vee \overline{x}_{2}) &  \\
\wedge & (\overline{x}_{3}\vee x_{2}) &  \\
\wedge & (x_{2}\vee & \overline{x}_{1}\vee x_{0}) \\
\wedge & (x_{2}\vee & x_{1}\vee \overline{x}_{0})%
\end{array}
.
\end{equation*}

A SSAT$\left( 2,3\right) $ is the three first clauses of $\varphi_2$, its solution is $%
\left[
\begin{array}{cc}
x_{3} & x_{2} \\
0 & 0%
\end{array}
\right].$

A SSAT$\left( 3,2\right) $ is the last two clauses of $\varphi_2$, its solutions are $%
\left[
\begin{array}{ccc}
x_{2} & x_{1} & x_{0} \\
1 & 0 & 0 \\
1 & 0 & 1 \\
1 & 1 & 0 \\
1 & 1 & 1 \\
0 & 0 & 0 \\
0 & 1 & 1%
\end{array}
\right] .$

Then SAT$\left( 4,7\right) $ has solution, because $0$ is the
common value for $x_{2}$. The solutions are $\left[
\begin{array}{cccc}
x_{3} & x_{2} & x_{1} & x_{0} \\
0 & 0 & 0 & 0 \\
0 & 0 & 1 & 1%
\end{array}
\right] .$

An example of the case 3) is the following $\varphi_3=$SAT$\left(
4,7\right) $
\begin{equation*}
\begin{array}{ccc}
\ \  & (x_{3}\vee \overline{x}_{2}) &  \\
\wedge & (\overline{x}_{3}\vee \overline{x}_{2}) &  \\
\wedge & (\overline{x}_{3}\vee x_{2}) &  \\
\wedge & (x_{2}\vee & \overline{x}_{1}\vee \overline{x}_{0}) \\
\wedge & (x_{2}\vee & \overline{x}_{1}\vee x_{0}) \\
\wedge & (x_{2}\vee & x_{1}\vee \overline{x}_{0}) \\
\wedge & (x_{2}\vee & x_{1}\vee x_{0})%
\end{array}
.
\end{equation*}

A SSAT$\left( 2,3\right) $ is the three first clauses of $\varphi_3$, its solution is $%
\left[
\begin{array}{cc}
x_{3} & x_{2} \\
0 & 0%
\end{array}
\right] .$

A SSAT$\left( 3,4\right) $ is the last four clauses of $\varphi_3$, its solutions are $%
\left[
\begin{array}{ccc}
x_{2} & x_{1} & x_{0} \\
1 & 0 & 0 \\
1 & 0 & 1 \\
1 & 1 & 0 \\
1 & 1 & 1%
\end{array}
\right].$

Then SAT$\left( 4,7\right) $ has not solution because there is not
a common value for the variable $x_{2}$.

%%%%%%%%%%%%%%%%%%%%%%%%%%%%%%%%%%%%%%%%%%%%%%%%%%%%%%%%%%%%%%%%%%%%%%%%%%%%
%%%%%%%%%%%%%%%%%%%%%%%%%%%%%%%%%%%%%%%%%%%%%%%%%%%%%%%%%%%%%%%%%%%%%%%%%%%%
%%%%%%%%%%%%%%%%%%%%%%%%%%%%%%%%%%%%%%%%%%%%%%%%%%%%%%%%%%%%%%%%%%%%%%%%%%%%
%%%%%%%%%%%%%%%%%%%%%%%%%%%%%%%%%%%%%%%%%%%%%%%%%%%%%%%%%%%%%%%%%%%%%%%%%%%%
\section{Properties for solving SAT}
~\label{sc:PropForSolSAT}
%%%%%%%%%%%%%%%%%%%%%%%%%%%%%%%%%%%%%%%%%%%%%%%%%%%%%%%%%%%%%%%%%%%%%%%%%%%%
%%%%%%%%%%%%%%%%%%%%%%%%%%%%%%%%%%%%%%%%%%%%%%%%%%%%%%%%%%%%%%%%%%%%%%%%%%%%
%%%%%%%%%%%%%%%%%%%%%%%%%%%%%%%%%%%%%%%%%%%%%%%%%%%%%%%%%%%%%%%%%%%%%%%%%%%%
%%%%%%%%%%%%%%%%%%%%%%%%%%%%%%%%%%%%%%%%%%%%%%%%%%%%%%%%%%%%%%%%%%%%%%%%%%%%

The translation of the SSAT$(n,m)$'s formulas can be arranged as a matrix $%
M_{n\times m}$ $=$ $\left[
\begin{array}{c}
b_{0} \\
\vdots \\
b_{m}%
\end{array}
\right] $ where each binary number $b_{i}$ corresponds to a each SSAT$(n,m)$%
's clause.

\begin{figure}[tbp]
\centerline{\psfig{figure=\IMAGESPATH/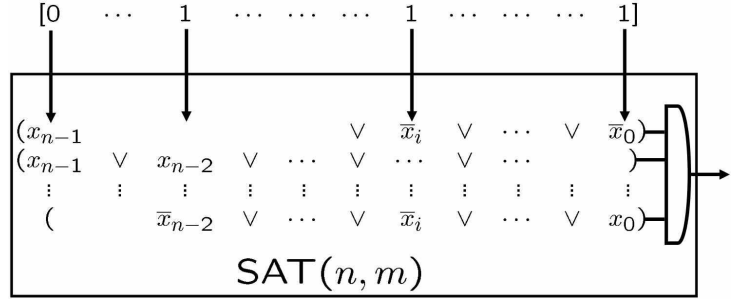, height=40mm}}~
\caption{ SAT$(n,m)$ is a white of box containing a circuit of logical
gates where each clause or row formula has the same number of boolean
variables.} \label{fig:BoxSATnxm}
\end{figure}

SSAT can be see as a logic circuit, it only depends of the
selection of the binary values assigned to $n$ lines, each line
inputs the corresponding binary value to its boolean variable
$x_i$. This is an important consideration because the complexity
of the evaluation as a circuit of a logic function is
$\mathbf{O}(k)$ with $k$ a fixed time. $k$ corresponds to the time
that the electrons activate the circuit in parallel lines for each
variable. Therefore, such evaluation can be considered as
$\mathbf{O}(1)$.  The figure~\ref{fig:BoxSATnxm} depicts
SAT$(n,m)$'s logic circuit.

The inner approach means to take the candidates from the translation of the
problem's formulas.

The outside approach means to the candidates from the problem's search space.

A deterministic approach means to take the candidates for solving a problem
with the data in its given order.

A probabilistic approach means to take the candidates for solving a problem
with no repetition and in random order, i.e., it likes a random permutation
of the numbers $[0,M].$

\begin{proposition}
~\label{prop:binNumBlock} Any disjunctive clause over the variables
$(x_{n-1},$ $\ldots,$ $x_0)$ corresponds a binary number $b$ $=$
$b_{n-1}b_n$ $\ldots$ $b_0 $.

\begin{enumerate}
\item $x \wedge \overline{x}\equiv 0$ where $\overline{x}$ is the complement
of $x$.

\item $b \wedge \overline{b} \equiv 0$ where $b$ and $\overline{b}$ correspond
to the translation of $x$ and $\overline{x}.$

\item $x(\overline{b}) \equiv 0$ and $\overline{x}(b) \equiv 0$
where the values of the boolean variables correspond to the bits
of $\overline{b}$ and $b.$
\end{enumerate}

\begin{proof}
Without loss of generality,  $x$ $=$ $x_{n-1}\vee
\overline{x}_{n-2}\vee \ldots \vee \overline{x}_0$ the translation
of $b$ and $\overline{x}$ $=$ $\overline{x}_{n-1}\vee x_{n-2}\vee
\ldots \vee x_0$ the translation of $\overline{b}$. Then $x \wedge
\overline{x}$ $=$ $(x_{n-1}\vee \overline{x}_{n-2}\vee \ldots \vee
\overline{x}_0)\wedge(\overline{x}_{n-1}\vee x_{n-2}\vee \ldots
\vee x_0)$ $\equiv$ $(x_{n-1}\wedge \overline{x}_{n-1}) \vee
(\overline{x}_{n-2} \wedge x_{n-2}) \vee \ldots \vee
(\overline{x}_0 \wedge x_0)$ $\equiv$ $0$.
\end{proof}
\end{proposition}

The translation of the SSAT's rows allows to define a table of binary
numbers. The following boards have not a satisfactory assignation in $\Sigma
$ and $\Sigma ^{2}$:

\begin{equation*}
\begin{tabular}{|c|}
\hline
$x_{1}$ \\ \hline
1 \\ \hline
0 \\ \hline
\end{tabular}
\ \ \
\begin{tabular}{|l|l|}
\hline
$x_{2}$ & $x_{1}$ \\ \hline
0 & 0 \\ \hline
1 & 1 \\ \hline
0 & 1 \\ \hline
1 & 0 \\ \hline
\end{tabular}%
\end{equation*}

I called unsatisfactory or blocked boards to the previous ones. It is clear
that they have not a solution because each binary number has its binary
complement. To find an unsatisfactory board is like to order the number and
its complement, by example: $000,$ $101,$ $110,$ $001,$ $010,$ $111,$ $011,$
and $100$ correspond to the unsatisfactory board:
\begin{equation*}
\begin{array}{c}
000 \\
111 \\
001 \\
110 \\
010 \\
101 \\
011 \\
100.%
\end{array}%
\end{equation*}

By inspection, it is possible to verify that the previous binary numbers
correspond to SSAT$(3,8)$ of the translated binary numbers with no solution
because any binary number is blocked by its complement binary number (see
prop.~\ref{prop:binNumBlock}). By example, $000$ and $111$ correspond to $%
(x_{2}\vee x_{1}\vee x_{0})\wedge (\overline{x}_{2}\vee \overline{x}_{1}\vee
\overline{x}_{0})$. Substituting by example $x_{2}=1,x_{1}=1,x_{0}=1$, we
get $(1\vee 1\vee 1)\wedge (0\vee 0\vee 0)$ $\equiv $ $(1)\wedge (0)$ $%
\equiv $ $0.$

\begin{proposition}
~\label{prop:SolSAT_binary} Let be SSAT$(n,m)$ with $m$ rows and $m<2^{n}$.
There is a satisfactory assignation that correspond to a binary string in $%
\Sigma^n$ as a number from $0$ to $2^{n}-1$.
\begin{proof}
Let  $s$ be any binary string that corresponds to a binary number
from $0$ to $2^n-1$, where $s$  has not its complement into the
translated formulas of the given SSAT$(n,m)$. Then $s$ coincide
with at least one binary digit of each binary number of the
translated rows formulas, the corresponding logical variable is 1.
Therefore, all rows are 1, i.e., $s$ makes SSAT$(n,m)$ = 1.
\end{proof}
\end{proposition}

The previous proposition point out when a solution $s\in[0, 2^n-1]$ exists
for SSAT. More important, SSAT can be see like the problem to look for a
number $s$ which its complements does not corresponded to the translated
numbers of the SSAT's formulas.

\begin{proposition}
~\label{prop:NoSolSAT_binary} Let be SSAT$(n,2^n)$ where its rows correspond
to the $0$ to $2^{n}-1$ binary numbers. Then it is an unsatisfactory board.
\begin{proof}
The binary strings of the values from $0$ to $2^n-1$ are all
possible assignation of values for the board. These strings
correspond to all combinations of $\Sigma^n$, and by the
prop.~\ref{prop:binNumBlock} SSAT$(n,2^n)$ has not solution.
\end{proof}
\end{proposition}

This proposition~\ref{prop:NoSolSAT_binary} states that if $m=2^n$ and SSAT
has different rows, then there is not a solution.

\begin{proposition}~\label{prop:sufcondSATo1}
For any SSAT$(n,m)$ the conditions a) $m$ rows formulas with $m < 2^n$
or b) $m$ different formulas with $m=2^n$ are sufficient conditions for
deciding the solution of any SSAT$(n,m)$ without further operations.
Under any of these two conditions the complexity for determining the
existence of the solution of SSAT is $\mathbf{O}(1).$
\begin{proof}
The results follow from the prop.~\ref{prop:SolSAT_binary} and
from the prop.~\ref{prop:NoSolSAT_binary}.
\end{proof}
\end{proposition}

\begin{proposition}
~\label{prop:NoSolSAT} Given SAT$(n,m).$ There is not solution, if
$L$ exists, where $L$ is any subset of boolean variables, with
their rows formulas isomorphic to an unsatisfactory board.
\begin{proof}
The subset $L$ satisfies the
proposition~\ref{prop:NoSolSAT_binary}. Therefore, it is not
possible to find satisfactory set of $n$ values for SAT$(n,m)$.
\end{proof}
\end{proposition}

Here, the last proposition depicts a condition to determine the no existence
of the solution for SAT. This property is easy to implement in an algorithm
using dynamic objects and when the set $L$ is associated with SSAT$(IV(L))$.
In order to determine that SAT contains a subproblem SSAT which it is an
unsatisfactory board, it is the same to verify that SSAT$(IV(L))$ has not
solution.

In similar way, the next proposition identifies when two SSAT can determine
the no solution of SAT.

\begin{proposition}
~\label{prop:NoSolSATby2SSAT} Given SAT$(n,m).$ There is not solution, if $%
L_1$ and $L_2$ exist, where $L_1$ and $L_2$ are any subset of
logical variables, such $L_1\cap L_2$ $\neq$ $\emptyset$, and the
unique solution of SSAT$(IV(L_1))$ and the unique solution of
SSAT$(IV(L_2))$ have no same values for the boolean variables in
$L_1\cap L_2$.
\begin{proof}
The result follows because the unique solution of SSAT$(IV(L_1))$
blocks the solution of SSAT$(IV(L_2))$ and reciprocally.
\end{proof}
\end{proposition}

\begin{proposition}
~\label{prop:SAT_TwoOne} Given SSAT$(n,m)$ as a circuit.

\begin{enumerate}
\item Let $k$ be the translation of any clause of SSAT$(n,m)$.

\item Let $k$ be any binary number, $k$ $\in$ $[0: 2^n-1]$.
\end{enumerate}

if SSAT${(n,m)}$$(k)=0$, then

\begin{enumerate}
\item SSAT$(n,m)(\overline{k})=0$ and the formulas of $k$ and $\overline{k}$
are in SSAT$(n,m).$

\item SSAT${(n,m)}$ $(k)=0$ and the translation of $\overline{k}$ is a
formula of SSAT$(n,m)$.
\end{enumerate}

\begin{proof}
\begin{enumerate}

\item When SSAT$(n,m)(k)=0$ is not satisfied, it is because the
formula of $\overline{k}$ is $0$.  SSAT${(n,m)}$ contains the
formulas of $k$ and $(\overline{k})=0$ (see
prop.~\ref{prop:binNumBlock}).

\item SSAT$(n,m)(k)=0$, then by prop.~\ref{prop:binNumBlock} the
translation of $\overline{k}$ is a formula of SSAT$(n,m)$.
\end{enumerate}

\end{proof}
\end{proposition}

\begin{proposition}
Let be $\Sigma={0,1}$ an alphabet. Given SSAT$(n,m)$, the set $\mathcal{S}$ $%
=$ $\{ x \in \Sigma^n |$ SSAT$(n,m)$$(x)$$=1$ $\}$ $\subset \Sigma^n$ of the
satisfactory assignations is a regular expression.
\begin{proof}
$\mathcal{S} \subset \Sigma^n$.
\end{proof}
\end{proposition}

The last proposition depicts that a set of binary strings $\mathcal{S}$ of
the satisfactory assignations can be computed by testing SSAT$(n,m)$$(x)$$=1$
for all $x\in[0,2^n-1]$, and the cost to determine $\mathcal{S}$ is $2^n$,
the number of different strings in $\Sigma^n$.

With $\mathcal{S}$ $\neq$ $\emptyset$ there is not opposition to accept that
SSAT$(n,m)$ has solution, no matters if $m$ is huge and the formulas are in
disorder or repeated. It is enough and sufficient to evaluate SSAT$(n,m)$$%
(x^\ast)$, $x^\ast \in \mathcal{S}$.

On the other hand, $\mathcal{S}$ $=$ $\emptyset$, there is not a direct
verification. This implies the condition $\mathcal{S}=\emptyset$ $%
\Leftrightarrow$ $\forall x\in \Sigma^n,$ SSAT$(n,m)(x)=0.$ It is necessary,
for verifying $\mathcal{S}$ $=$ $\emptyset$ to test all numbers in the
search space.

\begin{proposition}
~\label{prop:BuildSolSSAT}

{SSAT}$(n,m)$ \ has different row formulas, and $m \leq 2^{n}$. Any subset
of $\Sigma^n$ could be a solution for an appropriate {SSAT}$(n,m)$.
\begin{proof}
Any string of $\Sigma^n$ corresponds to a number in $[0,2^n-1]$
and a SSAT's formula.

 $\emptyset$ is the solution of a blocked board., i.e.,  for any
\SSAT \ with $m=2^n$.

For $m=2^n-1$, it is possible to build a \SSAT \  with only $x$ as
the solution. The blocked numbers $[0,2^n-1]$ $\setminus$ $\{x,
\overline{x}\}$ and $x$ are translated to SSAT's formulas. By
construction, \SSAT$(x)=1.$

 For $f$ different solutions. Let $x_1,\ldots,x_f$ be the given expected solutions.
 Build the set $C$ from the given solutions without any blocked
 pairs. Then the blocked numbers
$[0,2^n]$ $\setminus$ $\{y \in \Sigma^n |  x \in C, y=x \text{ or
} y= \overline{x}  \}$ and the numbers of $C$
 are translated to SSAT's rows.
\end{proof}
\end{proposition}

\begin{proposition}
~\label{prop:EvalMatchFixedPoint} Let be $y\in \Sigma ^{n}$, $%
y=y_{n-1}y_{n-2}\cdots y_{1}y_{0}.$ The following strategies of resolution
of SAT$(n,m)$ are equivalent.

\begin{enumerate}
\item The evaluation of SAT$(n,m)(y)$ as logic circuit.

\item ~\label{stp:match} A matching procedure that consists verifying that
each $y_{i}$ match at least one digit $s_{i}^{k}\in M_{n\times m},$ $\forall
k=1,\ldots ,m$.
\end{enumerate}

\begin{proof}
SAT$(n,m)(y)=1$, it means that at least one variable of each
clause is 1, i.e., each $y_i,$ $i=1,\ldots,n$ for at least one
bit, this matches to 1 in $s^k_j$, $k=1,\ldots, m$.
\end{proof}
\end{proposition}

% \begin{rem}
The evaluation strategies are equivalent but the computational cost is not.
The strategy~\ref{stp:match} implies at least $m \cdot n$ iterations. This
is a case for using each step of a cycle to analyze each variable in a clause
 or to count how many times a boolean variable is used. % \end{rem}

\begin{proposition}
An equivalent formulation of {SSAT}$(n,m)$\ is to look for a binary number $%
x^{\ast }$ from $0$ to $2^{n}-1.$

\begin{enumerate}
\item If $x^{\ast}\in \Sigma^n$ and $\overline{x}^{\ast }\notin M_{n\times
m} $ then SAT$(n,m)(x^{\ast })=1.$

\item If $m \leq 2^{n}-2$ and SSAT$(n,m)$ has different rows then $\exists
y^{\ast} \in [0, 2^{n}-1]$ and SSAT$(n,m)(y^{\ast })=1.$
\end{enumerate}

\begin{proof}  \

\begin{enumerate}
\item When $\overline{x}^{\ast }\notin M_{n\times m}$, this means
that the corresponding formula of $x^{\ast }$ is not blocked and
for each SAT$(n,m)$'s clause at least one boolean variable
coincides with one variable of $x^{\ast }.$ Therefore
SAT$(n,m)(x^{\ast })=1.$

\item $m \leq 2^{n}-2$, then $\exists y_1, y_2 \in[0, 2^{n}-1]$
with $y_1, y_2 \notin M_{n\times m}.$ There are two cases. 1)
$\overline{y}_2=y1$, therefore, SSAT$(n,m)(y_1)=1$. 2)
$\overline{y}_1 \in M_{n\times m}$, therefore
SSAT$(n,m)(\overline{y}_1)=1.$

\end{enumerate}
\end{proof}
\end{proposition}

This previous proposition, depicts the equivalence between SSAT with the
numerical problem to determine if there is a binary string, which is not
blocked by the binary translations of the SSAT's formulas. Only $k$ and its
complement $\overline{k}$ are opposed (see prop.~\ref{prop:binNumBlock}.
This point out the lack of other type of relations between the rows of SSAT.
More important, this proposition allows for verifying and getting a solution
for any SSAT$(n,m)$ without to evaluate SSAT$(n,m)$ as a function. By
example, SAT$(6,4)$ corresponds to the set $M_{6\times 4}$:

\begin{equation*}
\begin{tabular}{|l|l|l|l|l|l|l|}
\hline
& $x_{5}=0$ & $x_{4}=0$ & $x_{3}=0$ & $x_{2}=0$ & $x_{1}=0$ & $x_{0}=0$ \\
\hline
& $\overline{x}_{5}\vee $ & $\overline{x}_{4}\vee $ & $\overline{x}_{3}\vee $
& $\overline{x}_{2}\vee $ & $\overline{x}_{1}\vee $ & $\overline{x}_{0})$ \\
\hline
$\wedge ($ & $\overline{x}_{5}\vee $ & $\overline{x}_{4}\vee $ & $\overline{x%
}_{3}\vee $ & $\overline{x}_{2}\vee $ & $\overline{x}_{1}\vee $ & $x_{0})$
\\ \hline
$\wedge ($ & $x_{5}\vee $ & $x_{4}\vee $ & $x_{3}\vee $ & $x_{2}\vee $ & $%
x_{1}\vee $ & $\overline{x}_{0})$ \\ \hline
$\wedge ($ & $\overline{x}_{5}\vee $ & $x_{4}\vee $ & $x_{3}\vee $ & $%
\overline{x}_{2}\vee $ & $x_{1}\vee $ & $x_{0})$ \\ \hline
\end{tabular}
\text{ \ }
\end{equation*}

\begin{equation*}
\begin{tabular}{|l|l|l|l|l|l|}
\hline
$x_{5}$ & $x_{4}$ & $x_{3}$ & $x_{2}$ & $x_{1}$ & $x_{0}$ \\ \hline
$0$ & $0$ & $0$ & $0$ & $0$ & $0$ \\ \hline
$0$ & $0$ & $0$ & $0$ & $0$ & $1$ \\ \hline
$1$ & $1$ & $1$ & $1$ & $1$ & $0$ \\ \hline
$0$ & $1$ & $1$ & $0$ & $1$ & $1$ \\ \hline
\end{tabular}
\text{.}
\end{equation*}

How $y=000000\in\Sigma^6$ and $\overline{y}=111111 \notin M_{6\times 4}$
then SSAT$(n,m)(000000)=1.$ On the other hand, $y_1=100100 \in \Sigma^6$
with $\overline{y}_1=011011 \in M_{6\times 4}$ then SSAT$(n,m)(011011)=1.$

The next propositions, depicts the difficult for determining solving extreme
SSAT.

\begin{proposition}
~\label{prop:probsel} Let $n$ be large, and SSAT$(n,m)$ an extreme problem,
i.e., $|\mathcal{S}|$ $\leq 1$, and $m \gg 2^n$.

\begin{enumerate}
\item The probability for selecting a solution ($\mathcal{P}$$_{ss}(f)$)
after testing $f$ different candidates ($f<<2^n$) is $\approx 1 / 2^{2n}$
(it is insignificant).

\item Given $C$ $\subset $ $[0,2^{n}-1]$ with a polynomial cardinality,
i.e., $|C|$ $=$ $n^{k}$, with a constant $k>0.$ The probability that the
solution belongs $C$ ($\mathcal{P}$$_{s}(C)$) is insignificant, and more and
more insignificant when $n$ grows.

\item Solving SSAT$(n,m)$ is not efficient.
\end{enumerate}

\begin{proof} \

Assuming that $|\mathcal{S}|=1$.
\begin{enumerate}

\item The probability \Prob$_{ss}(f)$ corresponds to product of
the probabilities for be selected and be the solution.  For the
inner approach (i.e., the $f$ candidates are from the translations
of the SSAT$(n,m)$'s rows) \Prob$_{ss}(f)$ $=$ $1/\left( 2^{n}-2f
\right) \cdot 1 /2^n \approx 1/2^{2n} \approx 0.$ For the outside
approach (i.e., the $f$ candidates are from the $[0,2^n-1]$ the
search space) \Prob$_{ss}(f)$ $=$ $1/\left( 2^{n}-f \right) \cdot
1 /2^n \approx 1 /2^{2n} \approx 0.$

\item \Prob$(C)$ $=$ $n^k / 2^n.$ Then \Prob$_s(C)$ $=$ $n^k /
2^n$ $\cdot$ $1 /2^n$, and $lim_{n \to \infty} K n^k / 2^{n}$
(L'H\^{o}\-pi\-tal's rule) $=$ $0^+,$ $K>0.$ For $n$ large, $2^n-Kn^k
\approx 2^n,$ and $Kn^k \ll 2^n.$ Moreover, for the inner
approach, \Prob$_{ss}(n^k)$ $=$ $1/\left( 2^{n}-2n^k \right) \cdot
1 /2^n \approx 1/2^{2n} \approx 0.$ For the outside approach,
\Prob$_{ss}(n^k)$ $=$ $1/\left( 2^{n}-n^k \right) \cdot 1 /2^n
\approx 1 /2^{2n} \approx 0.$

 \item In any approach, inner or outside,
many rows of SSAT$(n,m)$ have large probability to be blocked,
because there is only one solution. Then the probability after $f$
iterations remains $1/ 2^{2n} \approx 0$. It is almost impossible
to find the solution with $f$ small or a polinomial number of $n$.
\end{enumerate}

Assuming that $|\mathcal{S}|=0$. \Prob$_s$ $=$ $0.$

\begin{enumerate}

\item[1,2] For the inner approach and for the outside approach,
\Prob$_{ss}(f)$ $=$ $0.$

\item[3] It is equivalent $\mathcal{S}=\emptyset$
$\Leftrightarrow$ SSAT$(n,m)(x)$
 $=$ $0,$ $\forall x \in [0,2^n-1].$ This means that it is
 necessary to test all the numbers in $[0,2^n-1].$

\end{enumerate}

\end{proof}
\end{proposition}

\begin{figure}[tbp]
\centerline{\psfig{figure=\IMAGESPATH/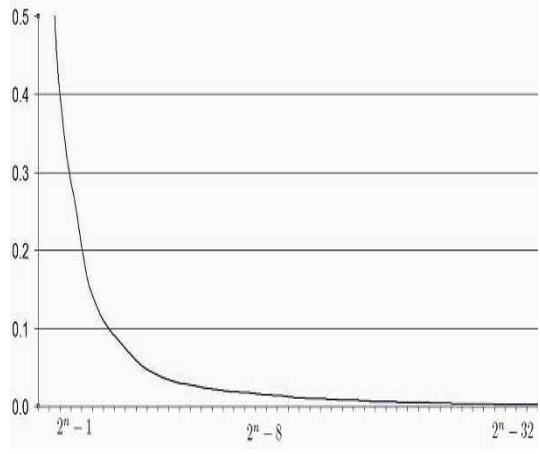,height=60mm,width=70mm
}}~
\caption{Behavior of the functions $P_e(t)$ and $P_i(t)$.}
\label{fig:probDecy}
\end{figure}

One important similarity between the extreme SSAT as a numerical problem
(see prop.~\ref{prop:BuildSolSSAT}) for one or none solution is the
interpretation to guest such type of solution. It is like a lottery but with
the possibility that there is not winner number. The exponential constant $%
2^n$ causes a rapidly decay as it depicted in fig.~\ref{fig:probDecy} where $%
t=2^n-1, 2^n-8, 2^n-32$.

\newpage
%%%%%%%%%%%%%%%%%%%%%%%%%%%%%%%%%%%%%%%%%%%%%%%%%%%%%%%%%%%%%%%%%%%%%%%%%%%%
%%%%%%%%%%%%%%%%%%%%%%%%%%%%%%%%%%%%%%%%%%%%%%%%%%%%%%%%%%%%%%%%%%%%%%%%%%%%
\section{Algorithms for SAT}
~\label{sc:algthmsSAT}
%%%%%%%%%%%%%%%%%%%%%%%%%%%%%%%%%%%%%%%%%%%%%%%%%%%%%%%%%%%%%%%%%%%%%%%%%%%%
%%%%%%%%%%%%%%%%%%%%%%%%%%%%%%%%%%%%%%%%%%%%%%%%%%%%%%%%%%%%%%%%%%%%%%%%%%%%
%%%%%%%%%%%%%%%%%%%%%%%%%%%%%%%%%%%%%%%%%%%%%%%%%%%%%%%%%%%%%%%%%%%%%%%%%%%%
%%%%%%%%%%%%%%%%%%%%%%%%%%%%%%%%%%%%%%%%%%%%%%%%%%%%%%%%%%%%%%%%%%%%%%%%%%%%

The previous sections depict characteristics and properties of SSAT.

The complexity for solving any SSAT$(n,m)$ has two cases:

\begin{enumerate}
\item It is $\mathbf{O(}1\mathbf{)}$ when $m<2^{n}.$

\item It is $\mathbf{O(}m\mathbf{)}\lessapprox \mathbf{O(2}^{n-1}\mathbf{)}$%
, when giving $m\geq 2^{n}.$ In this case, it is necessary at least one
carefully review of the SSAT's rows versus the search space $\left[ 0,2^{n}-1%
\right] $.
\end{enumerate}

The algorithms using the external approach are based in a random permutation
with only one lecture of the SSAT's rows or they stop if a satisfiable
assignation is found.

There are two source of data for solving SSAT$(n,m)$, 1) its $m$
rows or 2) the search space of all possible logical values for its
variables ($\Sigma ^{n}$). The second is large and $m$ could be
large also. Therefore, the efficient type of algorithms for
solving SSAT must be doing in one way without cycles, and with the
constraint that the total iterations must be related to $m<2^{n}$,
or $2^{n-1}$, or $2^{n}$. This is because the fixed point approach
or inside search(taking candidates from the translation SSAT's
formulas) and the outside approach or probabilistic approach
(taking candidates from the search space $[0,2^{n}-1].$

The situation for solving SSAT$(n,m)$ is subtle. Its number of rows could be
exponential, but for any SSAT$(n,m)$, there are no more than $2^{n}$
different rows, then $m\gg 2^{n}$ means duplicate rows. It is possible to
consider duplicate rows but this is not so important as to determine at
least one solution in $\Sigma ^{n}$. The search space $\Sigma ^{n}$
corresponds to a regular expression and it is easy to build by a finite
deterministic automata (Kleene's Theorem) but in order. However, to test the
binary numbers in order is not adequate.

Based in the algorithm~3 in ~\cite{arXiv:Barron2015b}, the update
information for any SSAT$(IV(r))$ is the following algorithm.

\begin{algorithm}
~\label{alg:UpdtSSAT}\textbf{Input:} List of object:
SSAT$(n^{\prime },m^{\prime })$, SSAT$(IV(r))$, $rw$: clause of
SAT$(n,m)$, where $r\subset X$.

\textbf{Output}: SSAT$(IV(r))$, $T:$ List of binary numbers such that, $x\in
T$, SSAT$(IV(r))$$(x)=1$.

\textbf{Variables in memory}: $T[0:2^{n-1}]$: list as an array of integers,
double link structure $previous$, $next$ : integer; $ct$:=$0$ : integer; $%
first=0$: integer; $last=2^{n}-1$: integer;

\vspace{-1mm}\noindent\hrulefill
\begin{easylist}
\renewcommand{\labelitemi}{\ }
 \setlength\itemsep{-0.2mm}
\item \textbf{if} SSAT$(IV(r))$ \textbf{does not exist then}

\item \hspace{0.5cm} create object SSAT$(IV(r))$

\item \textbf{end if}

\item with SSAT$(IV(r))$

\item \hspace{0.5cm} $k$ $=$ \textbf{Translate to binary formula} ($rw$);

\item \hspace{0.5cm} \textbf{if} $T[\overline{k}]$.previous \textbf{not
equal} $-1$ \textbf{\ or } $T[\overline{k}]$.next \textbf{not equal}
$-1$ \textbf{then}

\item \hspace{0.5cm} \hspace{0.5cm} // Update the links of $T$

\item \hspace{0.5cm} \qquad $T[T[\overline{k}]$.previous$]$.next $=T[%
\overline{k}]$.next;

\item \qquad \qquad  $T[T[\overline{k}]$.next$]$.previous $=T[%
\overline{k}]$.previous;

\item \qquad \qquad \textbf{if} $\overline{k}$ \textbf{equal} $first$
\textbf{then}

\item \qquad \qquad \qquad $first$ := $T[\overline{k}]$.next;

\item \qquad \qquad \textbf{end if}

\item \qquad \qquad \textbf{if} $\overline{k}$ \textbf{equal} $last$
\textbf{then}

\item \qquad \qquad \qquad $last$ := $T[\overline{k}]$.previous;

\item \qquad \qquad \textbf{end if}

\item \qquad \qquad $T[\overline{k}]$.next $=-1$;

\item \qquad \qquad  $T[\overline{k}]$.previous $=-1;$

\item \hspace{0.5cm} \qquad $ct:=ct+1;$

\item \hspace{0.5cm} \textbf{end if}

\item \hspace{0.5cm} \textbf{if} $ct$ \textbf{equal} $2^{n}$ \textbf{then}

\item \hspace{0.5cm} \hspace{0.5cm} \textbf{output:} There is not solution
for SAT$_{n\times m}$;

\item \hspace{0.5cm} \hspace{0.5cm} \textbf{everything stop};

\item \hspace{0.5cm} \textbf{end if}

\item \textbf{end with}

\item \textbf{return}
\end{easylist}
\vspace{-2mm}\noindent\hrulefill
\end{algorithm}

\begin{algorithm}
~\label{alg:SAT_one} \textbf{Input:} $n$, SAT$(n,\cdot)$.

\textbf{Output:} It determines if there is a solution or not of
SAT$(n,\cdot)$.

\textbf{Variables in memory}: $r$: set of variables of $X$; SSAT$(IV(r))$:
List of objects SSAT.

\vspace{-1mm}\noindent\hrulefill
\begin{easylist}
\renewcommand{\labelitemi}{\ }
 \setlength\itemsep{-0.2mm}
\item \textbf{While not} EOF(SAT$(n,\cdot)$);

\item \hspace{0.5cm} $r=$ SAT$(n,\cdot)$'s clause;

\item \hspace{0.5cm} // Update the information of SSAT$(IV(r))$

\item \hspace{0.5cm} Algorithm~\ref{alg:UpdtSSAT}(SSAT,r);

\item \textbf{end while};

\item \textbf{With the variables and the solutions of all}
SSAT$(IV(r))$ \textbf{determine the set} $\Theta$ \textbf{of the
corresponding} $\times \theta$ operation.

\item \textbf{if } $\Theta$ \textbf{is empty} \textbf{then}

\item \hspace{0.5cm} \textbf{Output}: "Algorithm~\ref{alg:SAT_one}. There is
not solution for the given SAT.

\hspace{0.5cm} The solutions of all SSAT$(IV(r))$ are incompatibles."

\item \hspace{0.5cm} \textbf{the process stops};

\item \textbf{else}

\item \hspace{0.5cm} \textbf{output}: "Algorithm~\ref{alg:SAT_one}. There is
solution for the given SAT.

\hspace{0.5cm} Let be $s$ any assignation $\in \Theta$. It is a
solution for SAT$(n.m)$, i.e.,

\hspace{0.5cm} All SSAT$(IV(r))$ are compatible."

\item \hspace{0.5cm} \textbf{the process stops};

\item \textbf{end if}

\vspace{-2mm}\noindent\hrulefill

\end{easylist}
\end{algorithm}

The next algorithm, is a version of the probabilistic algorithm 4 in~\cite%
{arXiv:Barron2015b}. It solves SAT$(n,m)$ in straight forward using an
outside approach, i.e, all the candidates are randomly and univocally
selected from search space $[0,2^n-1].$

\begin{algorithm}
~\label{alg:SAT_two} \textbf{Input:} $n$,  SAT$(n,\cdot)$.

\textbf{Output:} $s\in[0,2^n-1]$, such that SAT$(n,\cdot)$$(s)=1$
or SSAT has not solution (It determines if there is a solution or
not of SAT$(n,\cdot)$).

\textbf{Variables in memory}: $T[0: 2^{n-1}-1]$=$[0: 2^{n}-1]$: integer; $Mi$%
=$2^{n}-1$: integer; $rdm, a$: integer.

\vspace{-1mm}\noindent\hrulefill
\begin{easylist}
\renewcommand{\labelitemi}{\ }
 \setlength\itemsep{-0.2mm}
\item \textbf{for} i:=0 to $2^{n-1}-2$

\item \hspace{0.5cm} \textbf{if} $T[i]$ \textbf{equals} $i$ \textbf{then}

\hspace{0.5cm} \hspace{0.5cm} // \textbf{select randomly} $rdm \in [i+1,
2^{n-1}-1]$;

\item \hspace{0.5cm} \hspace{0.5cm} $rdm$ $=$ floor($rand()$ $\cdot$ $%
(Mi-i+1.5)$) + $(i+1)$;

\hspace{0.5cm} \hspace{0.5cm} // rand() return a random number in (0,1);

\hspace{0.5cm} \hspace{0.5cm} // floor(x) return the lower integer of x

\item \hspace{0.5cm}\hspace{0.5cm} $a$ $=$ $T[rdm]$;

\item \hspace{0.5cm}\hspace{0.5cm} $T[rdm]$ $=$ $T[i]$;

\item \hspace{0.5cm} \hspace{0.5cm} $T[i]$ $=$ $a$;

\item \hspace{0.5cm} \textbf{end if}

\item \hspace{0.5cm} $rdm$ = $0T[i]$;

\item \hspace{0.5cm} \textbf{if} SAT$(n,\cdot)(rdm)$ \textbf{equals} $1$ \textbf{%
then}

\item \hspace{0.5cm} \hspace{0.5cm} \textbf{output}: "Algorithm~\ref%
{alg:SAT_two}. There is solution for the given SAT.

\hspace{0.5cm} \hspace{0.5cm} The assignation $x$ is a solution for SAT$%
(n,m) $."

\item \hspace{0.5cm} \hspace{0.5cm} \textbf{the process stops};

\item \hspace{0.5cm} \textbf{end if}

\item \hspace{0.5cm} \textbf{if} SAT$(n,m)(\overline{rdm})$ \textbf{equals} $%
1$ \textbf{then}

\item \hspace{0.5cm} \hspace{0.5cm}\textbf{output}: "Algorithm ~\ref%
{alg:SAT_two}. There is solution for the given SAT.

\hspace{0.5cm} \hspace{0.5cm} The assignation $\overline{rdm}$ is
a solution for SAT$(n,\cdot)$."

\item \hspace{0.5cm} \hspace{0.5cm} \textbf{the process stops};

\item \hspace{0.5cm} \textbf{end if}

\item \textbf{end for}

\item $rdm=0T[2^{n-1}-1];$

\item \textbf{if} SAT$(n,m)(rdm)$ \textbf{equals} $1$ \textbf{then}

\item \hspace{0.5cm} \textbf{output}: "Algorithm~\ref{alg:SAT_two}. There is
solution for the given SAT.

\hspace{0.5cm} The assignation $rdm$ is a solution for
SAT$(n,\cdot)$."

\item \hspace{0.5cm} \textbf{the process stops};

\item \textbf{end if}

\item \textbf{if} SAT$(n,\cdot)(\overline{rdm})$ \textbf{equals} $1$ \textbf{then%
}

\item \hspace{0.5cm} \textbf{output}: "Algorithm~\ref{alg:SAT_two}. There is
solution for the given SAT.

\hspace{0.5cm} The assignation $\overline{rdm}$ is a solution for SAT$(n,\cdot)$%
."

\item \hspace{0.5cm} \textbf{the process stops};

\item \textbf{end if}

\item \textbf{output}: "Algorithm~\ref{alg:SAT_two}. There is not solution
for the given SAT.

Its rows cover all search space $[0,2^n]$, and they are blocked."

\item \textbf{the process stops};
\end{easylist}
\vspace{-2mm}\noindent\hrulefill
\end{algorithm}

The complexity of the previous algorithm is $\mathbf{O}\left(
2^{n-1}\right) $. No matters if the clauses of SAT$(n,\cdot)$ are
huge or duplicates or disordered, i.e., $\gg 2^{n}$.

\begin{algorithm}
~\label{alg:Parallel}\textbf{Input:} $\varphi=$SAT$(n,\cdot).$

\textbf{Output}: The solution of SAT$(n,\cdot).$

\textbf{Variables in memory}: List of object: SSAT$(n^{\prime
},m^{\prime })$.

\vspace{-1mm}\noindent\hrulefill
\begin{easylist}
\renewcommand{\labelitemi}{\ }
 \setlength\itemsep{-0.2mm}
\item \textbf{run in parallel}

\item \hspace{0.5cm} algorithm~\ref{alg:SAT_one}$(n,\varphi)$;

\item \hspace{0.5cm} algorithm~\ref{alg:SAT_two}$(n,\varphi)$;

\item \textbf{end run}
\end{easylist}
\vspace{-2mm}\noindent\hrulefill
\end{algorithm}

\begin{proposition}
~\label{prop:SSATforSAT} Let $n$ be large, and SAT$(n,\cdot)$,
$m\gg 2^{n}$.
Then SAT is solved at most $2^{n-1}$ steps by running algorithm~\ref%
{alg:Parallel}, which it runs in parallel the algorithms~\ref{alg:SAT_one}
and ~\ref{alg:SAT_two}.

\begin{proof}

Assuming enough time and memory. The two
algorithms~\ref{alg:SAT_one} and ~\ref{alg:SAT_two} run
independently in parallel.

The algorithm~\ref{alg:SAT_one} runs the algorithm~\ref{alg:UpdtSSAT}
with the current $r$ of SAT$(n,\cdot)$ until $m' = 2^{n'}$ or finishes
at the end of the SAT$(n,m)$'s rows. Where $n'=IV(r).n:$ number of
boolean variables, and $m'=IV(r).m':$ number of different rows. When
$m' = 2^{n'}$, the process stops because there is a blocked
SSAT$(IV(r))$. After finish to read the rows, with the variables and
the solutions of all SSAT$(IV(r))$, it determines from $\theta$ join
operation the set $\Theta$. if $\Theta$ $=$ $\emptyset$ then the
process stops, all SSAT$(IV(r))$ are incompatibles. Otherwise, $\Theta$
$\ne$ $\emptyset$ and any $s\in \Theta$ is a solution for
SAT$(n,\cdot).$

On the other hand, the algorithm ~\ref{alg:SAT_two} takes two
candidates at the same time $x=0T[k]$ and $\overline{x})$. If one
of them satisfies SAT$(n,\cdot)$ then stop. Otherwise, after all
candidates are tested, SAT$(n,m)$ has $2^n$ different rows then
the process stops because SAT$(n,m)$'s rows cover all search space
$[0,2^n]$, and they are blocked.

Finally, algorithm~\ref{alg:SAT_two} limits the steps at most
$2^{n-1}$ even if the number of clauses $\gg2^n.$
\end{proof}
\end{proposition}

%%%%%%%%%%%%%%%%%%%%%%%%%%%%%%%%%%%%%%%%%%%%%%%%%%%%%%%%%%%%%%%%%%%%%%%%%%%%
%%%%%%%%%%%%%%%%%%%%%%%%%%%%%%%%%%%%%%%%%%%%%%%%%%%%%%%%%%%%%%%%%%%%%%%%%%%%

\section{Complexity for SAT and SSAT}
~\label{sc:compleForSSAT}
%%%%%%%%%%%%%%%%%%%%%%%%%%%%%%%%%%%%%%%%%%%%%%%%%%%%%%%%%%%%%%%%%%%%%%%%%%%%
%%%%%%%%%%%%%%%%%%%%%%%%%%%%%%%%%%%%%%%%%%%%%%%%%%%%%%%%%%%%%%%%%%%%%%%%%%%%
%%%%%%%%%%%%%%%%%%%%%%%%%%%%%%%%%%%%%%%%%%%%%%%%%%%%%%%%%%%%%%%%%%%%%%%%%%%%
%%%%%%%%%%%%%%%%%%%%%%%%%%%%%%%%%%%%%%%%%%%%%%%%%%%%%%%%%%%%%%%%%%%%%%%%%%%%

An extreme SSAT is a problem with one solution or none but without
duplicates rows. For one solution, the simple comparison $m<2^n$ allows to
answer that the problem has a solution in one step. On the other hand, the
no solution case has complexity $\mathbf{O}(1)$, knowing that SSAT$(n,2^n)$
has different rows, there is nothing to look for. But again, to know that
SSAT$(n,2^n)$ has different rows, it has the cost of at least $\mathbf{O}%
(2^{n-1})$ by verifying at least one time the SSAT$(n,2^n)$'s rows
correspond to all combinations of $\Sigma^n$.

By example, the following SSAT$(3,7)$ has one solution $x_{2}=0$, $x_{1}=1$,
and $x_{0}=1$:
\begin{equation*}
\begin{tabular}{llccc}
&  &  & $\Sigma ^{3}$ & $[0,7]$ \\
& $(\overline{x}_{2}\vee \overline{x}_{1}\vee \overline{x}_{0})$ &  & 000 & 0
\\
$\wedge $ & $(\overline{x}_{2}\vee \overline{x}_{1}\vee x_{0})$ &  & 001 & 1
\\
$\wedge $ & $(\overline{x}_{2}\vee x_{1}\vee \overline{x}_{0})$ &  & 010 & 2
\\
$\wedge $ & $(\overline{x}_{2}\vee x_{1}\vee x_{0})$ &  & 011 & 3 \\
$\wedge $ & $(x_{2}\vee \overline{x}_{1}\vee x_{0})$ &  & 101 & 5 \\
$\wedge $ & $(x_{2}\vee x_{1}\vee \overline{x}_{0})$ &  & 110 & 6 \\
$\wedge $ & $(x_{2}\vee x_{1}\vee x_{0})$ &  & 111 & 7%
\end{tabular}
\text{ \ }
\end{equation*}

By construction, the unique solution is the binary string of $3$. It
corresponds to the translation $(\overline{x}_{2}\vee x_{1}\vee x_{0})$. It
satisfies SSAT$(3,7)$, as the assignation $x_{2}=0$, $x_{1}=1$, and $x_{0}=1$%
. It is not blocked by $100$, which corresponds to the missing formula $%
(x_{2}\vee \overline{x}_{1}\vee \overline{x}_{0})$ (The complement of the
formula $3$). The other numbers $0,1,2$ are blocked by $5,6,7$.

An extreme SSAT has the next relation with a SAT:

\begin{enumerate}
\item The unique solution of SSAT$\left( n,2^{n}-1\right) $
corresponds to a SAT with $n$ rows, where each row corresponds to
each variable of the
solution. By example, for the previous SSAT$(3,7)$ its corresponding SAT$%
\left( n,n\right) $ is $%
\begin{array}{r}
\ \ \left( \overline{x}_{2}\right) \\
\wedge \left( x_{1}\right) \\
\wedge \left( x_{0}\right)%
\end{array}
.$

\item The no solution case SSAT$\left( n,m\right) $ corresponds to a
blocked board, by example, SAT$\left( 1,2\right) $ could be $%
\begin{array}{r}
\left( \overline{x}_{0}\right) \\
\wedge \left( x_{0}\right)%
\end{array}
.$
\end{enumerate}

The extreme SSAT problem is designed to test how difficult is to
determine one or none solution knowing only $n$ the number of
variables, and $m$ the number of rows. It is extreme because $m
\gg 2^n$ could be huge. However, the corresponding versions of the
extreme SSAT have two easy SAT problems. It is nor complicated but
laborious to verify or build both SSAT and SAT. The next
proposition allows to transform SAT in SSAT, and reciprocally.

\begin{proposition}
~\label{prop:FtFvyFnv} Let $F$ be a boolean formula and $v$ a
boolean variable, which is not in $F$. Then
\begin{equation*}
\begin{array}{ccc}
\left( F \right) & \equiv &
\begin{array}{cc}
& \left( F \vee v\right) \\
\wedge & \left( F\vee \overline{v}\right)%
\end{array}%
\end{array}%
\end{equation*}

\begin{proof}
The result follows from factorization and distribution laws: \[
\begin{array}{ccc}
\left( F \right)  & \equiv  & \left( F \wedge \left( v\vee
\overline{v}\right)
\right)  \end{array}%
\]
.
\end{proof}
\end{proposition}

\begin{figure}[tbp]
\centerline{\psfig{figure=\IMAGESPATH/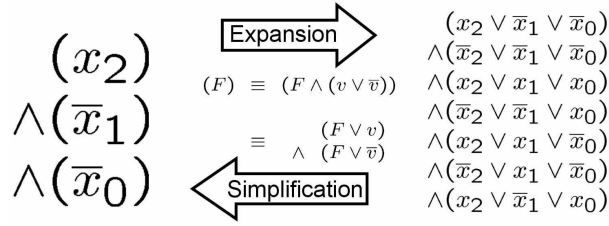, height=30mm}}~
\caption{The relationship between SAT and SSAT.} \label{fig:SATtSSAT}
\end{figure}

The previous proposition allows to define the operations expansion
and simplification (see fig.~\ref{fig:SATtSSAT}):

\begin{enumerate}
\item Expansion. Add the two corresponding clauses for each
boolean variable, which are not in $F$, where $F$ is a factor or
part of a boolean formula.

\item Simplification. Two clauses simplifies into one clause by
the factorization: $\left( F \vee v \right) \wedge \left( F \vee
\overline{v}\right) \equiv \left( F \right).$
\end{enumerate}

\begin{proposition}
~\label{prop:SATtSSAT} SSAT and SAT are equivalent, i.e., any SSAT can be
transformed in SAT, and reciprocally by using by prop.~\ref{prop:FtFvyFnv}.
\begin{proof}
Giving a SAT, the formulas are completed by expansion (see
fig.~\ref{fig:SATtSSAT}) from the previous proposition to build a
SSAT. Reciprocally, Given a SSAT by using factorization it could
be simplified to a SAT or there is nothing to do. In any case, SAT
and its expansion SSAT or SSAT and its simplified SAT has the same
set of solution by prop.~\ref{prop:FtFvyFnv}.
\end{proof}
\end{proposition}

It is important to note, that for solving SSAT the complexity is
bounded quasi lineal as a function of the number of SSAT's rows
and it is bounded $2^n$, because the size of the search space of
all possible solutions $\Sigma^n$ or $[0,2^n-1]$ but in function
of the $n$ the number of the logical variables of the problem.
More important, it is trivial to solve SSAT when $m<2^n$ and when
$m=2^n$ and SSAT$(n,2^n)$'s rows are different.

On the other hand, when $m\gg 2^n$ it is by construction that at
least one checking between SSAT and its search space is necessary
to determine its solution. This up an objection to disqualify the
extreme problem because it is by construction exponential in the
number of SSAT's rows. Moreover, what could be a source of such
type of problem or it is a theoretical curiosity. In my opinion,
it is not a curiosity but a future technological issue.
In~\cite{arXiv:Barron2015b}, the algorithm 5 is an hybrid
hardware-software over quantum computation and a SAT as an
appropriate electronic logical circuit. The creation of novel
electronic circuits is near to the level of the crystalline
structures. This means that figure~\ref{fig:BoxSATnxm} could
correspond to a crystalline structure. Here, for technical reasons
existence and solutions are necessary to determine.

On the other hand, SAT as the general problem could have rows with
any length and in any order. It is trivial to build a random SAT
generator problems. The minimum parameter is $n$ the number of
logical variables, and the output are $m$ the numbers of rows and
the rows. Solving an arbitrary SAT by using algebra increase the
complexity because it requires to compare and match rows and
variables in some special order and with ad-hoc and appropriate
structure for finding factor or parts where to apply the
operations expansion or simplification (see
fig.~\ref{fig:SATtSSAT}). This has as consequence more than one
lecture or access of the SAT's rows, i.e., more than $m$
operations, for ordering and matching SAT's rows and variables
which it is not appropriate when $m$ is large, i.e., $m\geq 2^{n}$
or $m\gg 2^{n}.$

The algorithm~\ref{alg:Parallel} and particularly the
algorithm~\ref{alg:SAT_one} do one lecture of the SAT's rows for
extracting information of its subproblem SSAT. Together with the
$\theta$ operation at the step 9, the algorithm~\ref{alg:SAT_one}
determines if such SSAT are compatibles or not. The iterations are
less than $m$ because the detection of a blocked SSAT is a
sufficient condition to determine the no solution of the given
SAT. On the other branch, the algorithm~\ref{alg:SAT_two} divides
the search space in two sections for testing two candidates at the
same time in one iteration. It is possible to divide the search
space in more sections but the candidates for testing in each step
grows exponentially, 4,8, $2^k$, $\ldots$. The testing of the
candidates can be in parallel and the complexity can be reduced to
$2^{n-k}$ where $2^k$ are the candidates for testing. In our case,
$2^1=2$, therefore the search space is exploring in $2^{n-1}$
steps. This means that it is necessary to have $2^k$ processors
for testing $2^k$ candidates to get a lower upper bound of
$2^{n-k}$ iterations. Taking in consideration that $2^k$ processor
with $k \gg 0$ is not posible, the lower upper bound is $2^{n-1}$.

The algorithm~\ref{alg:SAT_one} is capable to process under enough
memory and time with complexity $\mathbf{O}\left(|\varphi|+
\text{time}(\times \theta \, \forall \, \text{SSAT}(IV(r)) \right)$ any
$\varphi=r,s$-SAT. The $r,s$-SAT formulation means formulas in CNF with
clauses of $r$ variables where any variable is repeated at most $s$
times. In particular, any $r,1$-SAT or $r,2$-SAT or $r,r$-SAT can be
solved as the algorithm of the state of
art~\cite{Pudlak1998,Zhang:2001:ECD:603095.603153,
Zhang2002,TOVEY198485}. Considering that $\text{time}(\times \theta \,
\forall \, \text{SSAT}(IV(r))$ is solved by short-cut strategies. To
detect $r,1$-SAT is when $\bigcap_{r\in \varphi} IV(r)=\emptyset$, and
the solutions(SSAT$(IV(r))$) $\neq$ $\emptyset$. It is trivial and fast
to create any $s\in$ $\times$ solutions(SSAT$(IV(r))$). It is similar
for $r,2$-SAT. Finally,
\begin{proposition}
For any $\varphi = r,r$-SAT with exactly $r$ clauses. The complexity to
determine $\varphi \in$ SAT is $\mathbf{O}\left(1\right)$.
\begin{proof}
By constructing, $\varphi$, it is a SSAT$(r,r)$, i.e., its parameters
$r,r$ are giving. The result follows by the
proposition~\ref{prop:sufcondSATo1}.
\end{proof}
\end{proposition}

%%%%%%%%%%%%%%%%%%%%%%%%%%%%%%%%%%%%%%%%%%%%%%%%%%%%%%%%%%%%%%%%%%%%%%%%%%%%
%%%%%%%%%%%%%%%%%%%%%%%%%%%%%%%%%%%%%%%%%%%%%%%%%%%%%%%%%%%%%%%%%%%%%%%%%%%%

\section*{Conclusions and future work}
~\label{sc:conclusions and future work}
%%%%%%%%%%%%%%%%%%%%%%%%%%%%%%%%%%%%%%%%%%%%%%%%%%%%%%%%%%%%%%%%%%%%%%%%%%%%
%%%%%%%%%%%%%%%%%%%%%%%%%%%%%%%%%%%%%%%%%%%%%%%%%%%%%%%%%%%%%%%%%%%%%%%%%%%%
%%%%%%%%%%%%%%%%%%%%%%%%%%%%%%%%%%%%%%%%%%%%%%%%%%%%%%%%%%%%%%%%%%%%%%%%%%%%
%%%%%%%%%%%%%%%%%%%%%%%%%%%%%%%%%%%%%%%%%%%%%%%%%%%%%%%%%%%%%%%%%%%%%%%%%%%%

The results here confirm that there is a upper limit for the SAT's
complexity (It was predicted in the
article~\cite{arXiv:Barron2015b}). The outside approach and the
evaluation of SSAT as a circuit correspond to the probabilistic
type of method allow to build the stable
algorithm~\ref{alg:SAT_two}. This algorithm
is a more detailed version of the probabilistic algorithm 4 of ~\cite%
{arXiv:Barron2015b}. It states the upper bound $2^{n-1}$ steps for solving
any SAT.

The main result is the impossibility to build an efficient
algorithm for solving Extreme SAT because the steps grows $\approx 2^n$ with $n$ a big number of variables.

The consequences of the no existence of an
efficient algorithm for an Extreme SAT, open the question does the Extreme SAT can be built but without the possibility to solve it? It seems that an Extreme SAT can be used to keep a secret number without the possibility to break it even the most advanced Numerical intensive computers or supercomputers. Nevertheless quantum computation and the
hybrid hardware-software will open a new era in the computer
science.

%%%%%%%%%%%%%%%%%%%%%%%%%%%%%%%%%%%%%%%%%%%%%%%%%%%%%%%%%%%%%%%%%%%%%%%%%%%%
%%%%%%%%%%%%%%%%%%%%%%%%%%%%%%%%%%%%%%%%%%%%%%%%%%%%%%%%%%%%%%%%%%%%%%%%%%%%
% \bibliographystyle{abbrv}
% \bibliography{\BIBPATH/NPComplexity_v15}

\end{document}